\documentclass[12pt]{iopart}
\usepackage{graphicx}
\usepackage{hyperref}

\begin{document}
\title[A class of permanent magnetic lattices for ultracold atoms]{A class of permanent
magnetic lattices for ultracold atoms}
\author{Saeed Ghanbari, Tien D Kieu and Peter Hannaford
}
\address{Centre for Atom Optics and Ultrafast Spectroscopy \\
and  ARC Centre of Excellence for Quantum Atom Optics, \\
Swinburne University of Technology, Melbourne, Australia 3122}
\begin{abstract}
We report on a class of configurations of permanent magnets on an
atom chip for producing 1D and 2D periodic arrays of magnetic
microtraps with non-zero potential minima and variable barrier
height for trapping and manipulating ultracold atoms and quantum
degenerate gases. We present analytical expressions for the relevant
physical quantities and compare them with our numerical results and
with some previous numerical calculations. In one of the
configurations of permanent magnets, we show how it is possible by
changing  the angle between the crossed periodic arrays of magnets
to go from a 1D array of 2D microtraps to a 2D array of 3D
microtraps and thus to continuously vary the barrier heights between
the microtraps. This suggests the possibility of performing a type
of `mechanical' BEC to Mott insulator quantum phase transition in a
magnetic lattice. We also discuss a configuration of magnets which
could realize a two-qubit quantum gate in a magnetic lattice.
\end{abstract}

\section{Introduction}
Magnetic lattices consisting of periodic arrays of current-carrying
wires~\cite{Yin,Grabowski} or permanent magnetic
films~\cite{Hinds,Ghanbari} have recently been proposed as  an
alternative approach to optical lattices~\cite{Jaksch,Greiner} for
trapping and manipulating small clouds of ultracold atoms and
quantum degenerate gases, including Bose-Einstein condensates.
 Magnetic lattices may be considered as complementary to optical
lattices, in much the same way as magnetic traps are complementary
to optical dipole traps: they do not require (intense and stable)
laser beams and there is no decoherence or light scattering due to
spontaneous emission; they can produce highly stable and
reproducible potential wells leading to high trap frequencies; and
only atoms in low magnetic field seeking states are trapped, thus
allowing the possibility  of performing rf evaporative cooling $in$
$ situ$ in the lattice and the study of very low temperature
phenomena in a periodic lattice. Simple 1D magnetic lattices
consisting of periodic arrays of traps or waveguides have been
constructed using both current carrying wires~\cite{Gunther,Hansel}
and permanent
magnets~\cite{Barb,Sinclair,SidorovMcSch,Boyd} on atom chips.\\

\noindent In a recent paper, we proposed  a 2D magnetic lattice
consisting of two crossed layers of periodic arrays of parallel
rectangular magnets~\cite{Ghanbari}. In this paper, we report on a
new class of configurations of permanent magnets comprising $four$
periodic arrays of square magnets of different thickness for
producing 1D and 2D arrays of magnetic microtraps with non-zero
potential minima and variable barrier height.  Analytical
expressions are presented for various physical quantities and
compared with numerical calculations.  In one of the configurations
of magnets, we show that by varying the angle between the crossed
arrays of magnets it is possible to go from a 1D array of microtraps
to a 2D array of microtraps.  The 2D magnetic lattices consisting of
four periodic arrays of square magnets of different thickness
reported here may prove easier to implement experimentally than the
crossed  periodic arrays of parallel rectangular magnets proposed
previously~\cite{Ghanbari}.
\section{Periodic arrays of square permanent magnets with bias
magnetic fields }\label{sec2} 
\begin{figure}[tbp]
\begin{center}
$\begin{array}{cc}
\includegraphics[angle=0,width=6cm]{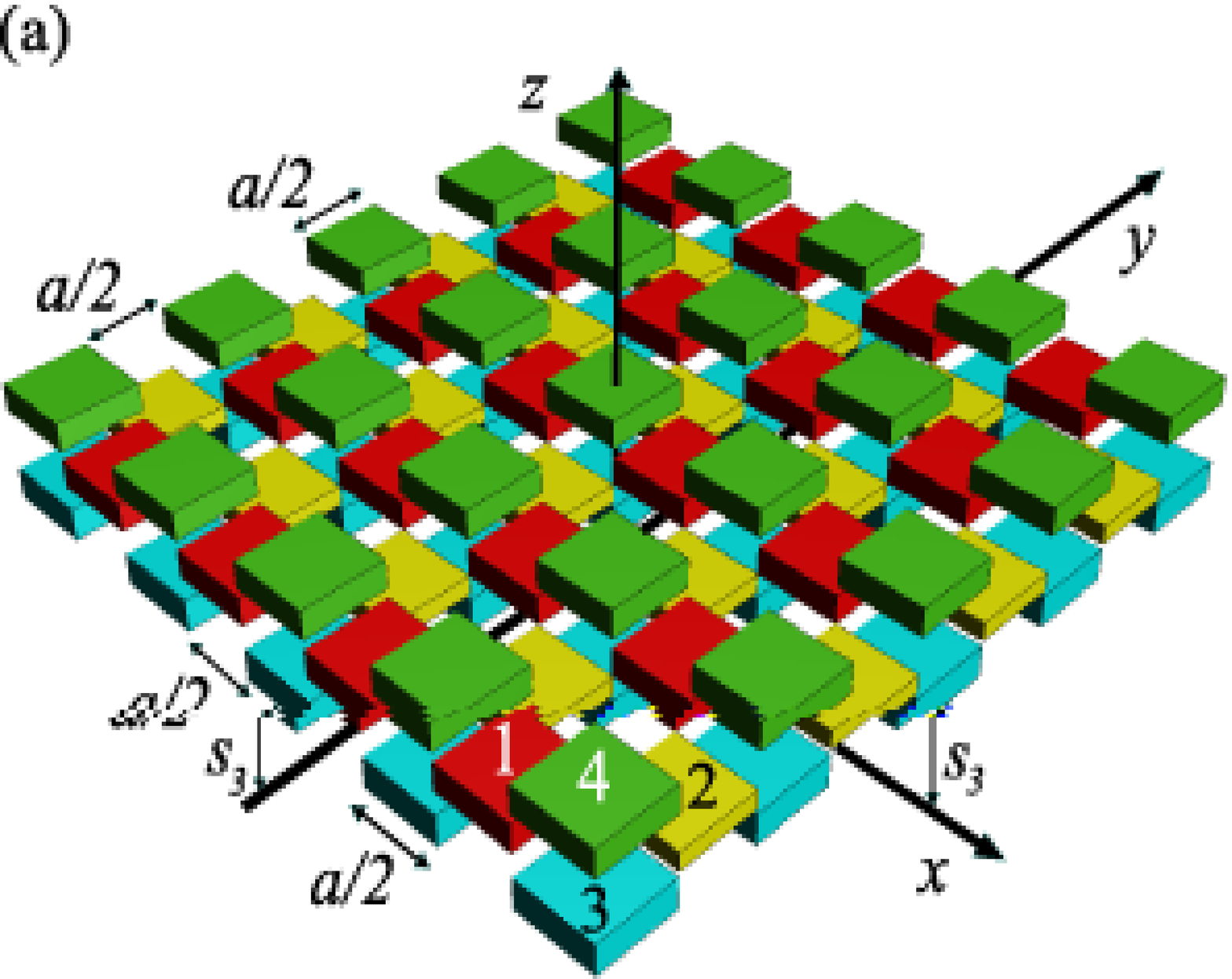}  &
\includegraphics[angle=0,width=6cm]{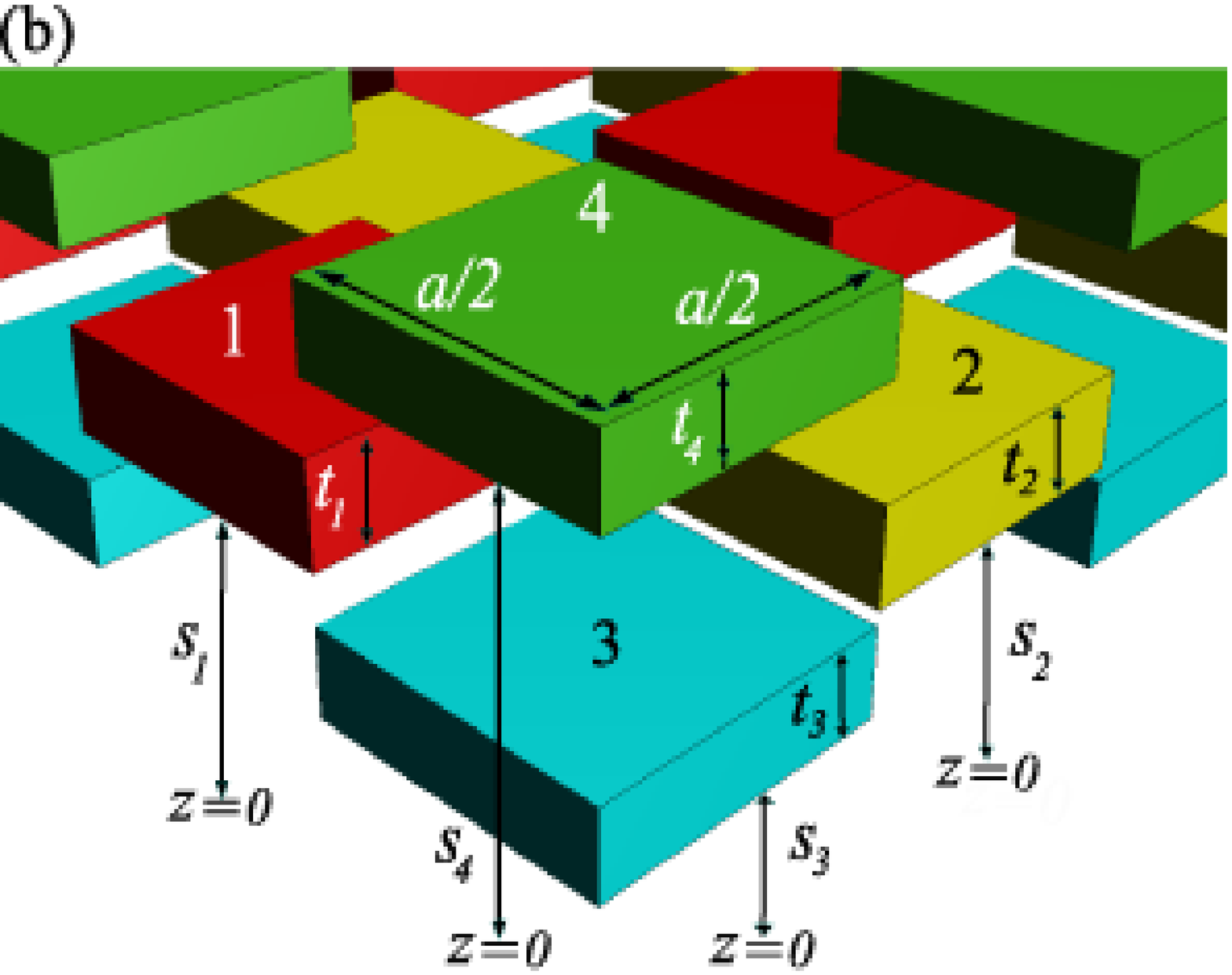}\\
\includegraphics[angle=0,width=6cm]{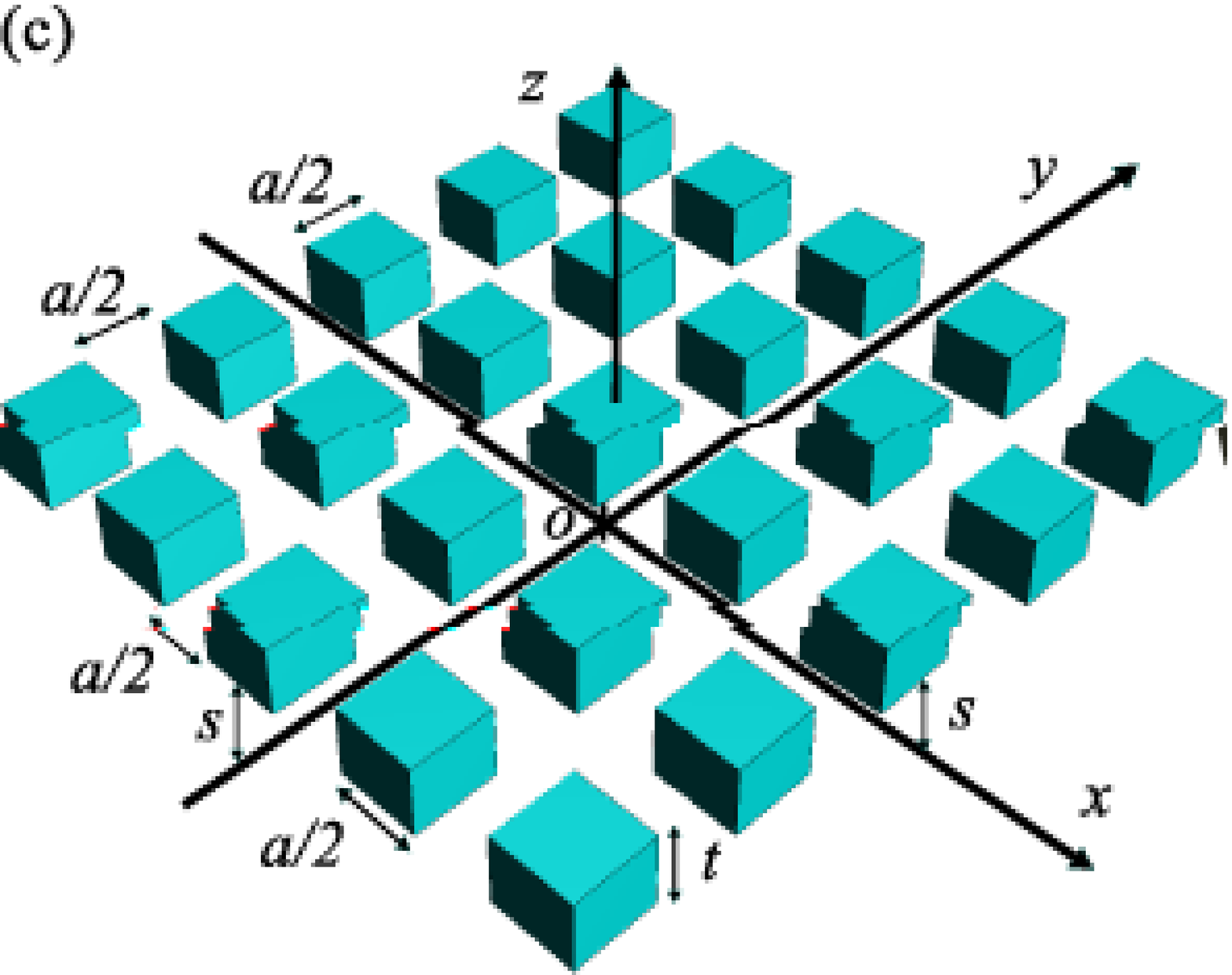}  &
\includegraphics[angle=0,width=6cm]{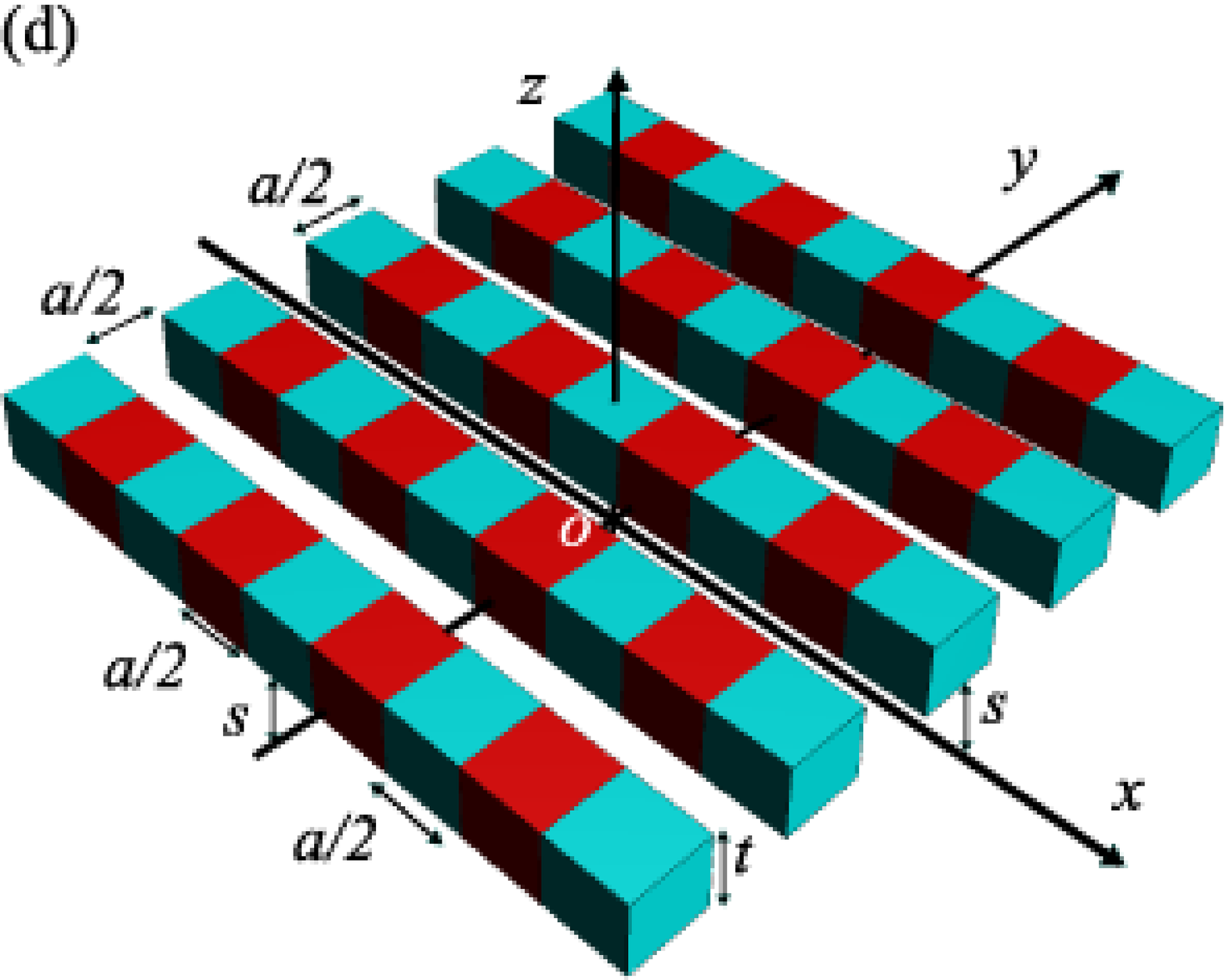}
\end{array}$
\caption{(a) and (b) Four periodic arrays of square permanent
magnets with thicknesses $t_1$, $t_2$, $t_3$ and $t_4$,
respectively. (c) Array of square permanent magnets of thickness
$t$, at a distance $s$ from the plane $z=0$, with periodicity $a$
along the $x$- and $y$-directions and perpendicular magnetization
$M_z$. (d) Single periodic array of parallel, rectangular magnets
with perpendicular magnetization.}\label{figure1}
\end{center}
\end{figure}

\Fref{figure1}(a)-(b) shows a configuration of four periodic arrays
of square
 magnetic slabs with thicknesses $t_1$, $t_2$, $t_3$ and
$t_4$, respectively.  In \sref{sec One array of square}, we find the
components of the magnetic field due to a single array of square
magnetic slabs and then use our results to obtain the total magnetic
field due to the four arrays of   square magnetic slabs and an
external bias field ${\bf B_1}=B_{1x}{\hat x}+B_{1y}{\hat y}$.

\subsection{Single periodic array of square magnets}\label{sec One array of square}
Here, we consider a single infinite periodic array of square
magnetic slabs of thickness $t$, at a distance $s$ from the plane
$z=0$, with periodicity $a$ along the $x$- and $y$-directions and
perpendicular magnetization $M_z$ [\fref{figure1}(c)]. The sum of
this configuration and a similar one displaced by ${a/2}$ along the
$x$-axis gives an infinite periodic array of parallel rectangular
magnets with periodicity $a$ along the $y$-direction and the same
magnetization $M_z$ and thickness $t$
 [\fref{figure1}(d)]. For distances
from the surface which are large compared with $a/2\pi$, the
components of the magnetic field are~\cite{Ghanbari} \numparts
\begin{equation}\label{E1a}
B_{x} = 0
\end{equation}
\begin{equation}\label{E1b}
B_{y} = B_{0y} \sin (ky) {\rm e}^{-kz}
\end{equation}
\begin{equation}\label{E1c}
B_{z} =  B_{0y} \cos (ky) {\rm e}^{-kz}
\end{equation}
\endnumparts
where $ B_{0y} = B_{0}( {\rm e}^{kt}-1){\rm e}^{ks}$, $k={2\pi/a}$
and  $B_0=4M_z$ (Gaussian units). Considering the corresponding
surface currents around the magnetic slabs in \fref{figure1}(d), we
find that the array of blue (grey in grey style) square magnets
produces the same component of magnetic field along the $y$-axis as
the array of red (black in grey style) magnets. Thus, the
$y$-component for the array of blue (or red) magnets can be written
as half that of \eref{E1b}. Furthermore, \fref{figure1}(c) is
symmetrical with respect to $x$ and $y$, and therefore,  for this
infinite array of square  magnets, the dependence of the
$x$-component of the magnetic field on $x$ is the same as the
dependence of  the $y$-component of the magnetic field on $y$.  On
the other hand, the $z$-component of the magnetic field should be
symmetrical with respect to $x$ and $y$ and also needs to be maximum
at $x=y=0$. Thus the components of the magnetic field for the
configuration of magnets in \fref{figure1}(c) may be written as
\numparts
\begin{equation}\label{E2a}
B_{x} = B_{01} \sin (kx) {\rm e}^{-kz}
\end{equation}
\begin{equation}\label{E2b}
B_{y} = B_{01} \sin (ky) {\rm e}^{-kz}
\end{equation}
\begin{equation}\label{E2c}
B_{z} =  B_{01} [\cos (kx)+\cos (ky)] {\rm e}^{-kz}
\end{equation}
\endnumparts
where $z \gg {a\over{2\pi}}+ s + t $ and $ B_{01} = {B_{0}\over 2}(
{\rm e}^{kt}-1){\rm e}^{ks}$.
\subsection{Four periodic arrays of square magnets}\label{sec Four array of square}
Using \eref{E2a}-\eref{E2c}, we can write the components of the
magnetic field due to the four arrays of square permanent magnetic
slabs [\fref{figure1}(a)-(b)] with external bias  field ${\bf
B_1}=B_{1x}{\hat x}+B_{1y}{\hat y}$ as \numparts
\begin{equation}\label{E3a}
B_{x} = B_{0x} \sin (kx) {\rm e}^{-kz}+B_{1x}
\end{equation}
\begin{equation}\label{E3b}
B_{y} = B_{0y} \sin (ky) {\rm e}^{-kz}+B_{1y}
\end{equation}
\begin{equation}\label{E3c}
B_{z} =  [B_{0x} \cos (kx)+B_{0y} \cos (ky)] {\rm e}^{-kz}
\end{equation}
\endnumparts
where $z \gg {a\over{2\pi}}+ max(s_1+t_1, s_2+t_2, s_3+t_3,
s_4+t_4)$ and \numparts
\begin{equation}\label{E4a}
B_{0x} = -B_{01}+ B_{02}+B_{03}+B_{04}
\end{equation}
\begin{equation}\label{E4b}
B_{0y} = B_{01}-B_{02}+B_{03}+B_{04}
\end{equation}
\begin{equation}\label{E4c}
B_{0i} = {B_{0}\over 2}( {\rm e}^{kt_i}-1){\rm e}^{ks_i}, \qquad
(i=1, 2, 3, 4)
\end{equation}
\endnumparts
The magnitude of the magnetic field above the  magnetic arrays is
then
\begin{eqnarray}
 B(x,y,z) &=& \Bigl\{B_{1x}^2+B_{1y}^2  \nonumber \\
 &+& 2 [B_{0x}B_{1x} \sin(kx) + B_{0y}B_{1y} \sin(ky)]{\rm
e}^{-kz}  \nonumber \\
&+& [B_{0x}^2+B_{0y}^2 + 2 B_{0x}B_{0y} \cos(kx)\cos(ky) ]{\rm
e}^{-2kz} \Bigr\}^{1\over 2} \label{E5}
\end{eqnarray}
This configuration of square  magnets gives a 2D periodic lattice of
magnetic traps with {\it non-zero} potential minima given by
\begin{equation}\label{E6}
B_{min}=  \frac{|B_{0x}B_{1y}  - B_{0y}B_{1x}|
}{({B_{0x}^2+B_{0y}^2})^{1\over 2}}
\end{equation}
which are located at \numparts
\begin{equation}\label{E7a}
x_{min}=\left(n_x+{1 \over 4}\right) a,  \qquad n_x=0,\pm 1,\pm
2,\cdots
\end{equation}
\begin{equation}\label{E7b}
y_{min}=\left(n_y+{1 \over 4}\right) a,  \qquad  n_y=0,\pm 1,\pm
2,\cdots
\end{equation}
\begin{equation}\label{E7c}
z_{min}={a \over {2\pi}} \ln \left({{ B_{0x}^2+B_{0y}^2}\over
-B_{0x}B_{1x}-B_{0y}B_{1y}}\right)
\end{equation}
\endnumparts
If we require  a 2D lattice with {\it non-zero} magnetic field
minima, \eref{E6} and \eref{E7c} impose the  constraints
 \numparts
\begin{equation}\label{E8a}
{B_{0x} B_{1y}} \neq {B_{0y} B_{1x}}
\end{equation}
\begin{equation}\label{E8b}
{ B_{0x}^2+B_{0y}^2}>-B_{0x}B_{1x}-B_{0y}B_{1y} >0
\end{equation}
\endnumparts
on the geometrical parameters and the components of the bias
magnetic field, where according to \eref{E4a}-\eref{E4c}, $B_{0x}$
and $B_{0y}$ are independent of the magnetization and depend only on
the geometrical constants $t_1, t_2, t_3, t_4, s_1, s_2, s_3$ and
$s_4$.
\\

\noindent The potential barrier heights in the three directions are
given by \numparts
\begin{equation}\label{E9a}
\Delta B^{x_j} =  \left[\frac{4
B_{0{x_j}}^2B_{1{x_j}}^2+(B_{0x}B_{1y} +B_{0y}B_{1x})^2
}{{B_{0x}^2+B_{0y}^2}}\right]^{1\over 2} \hspace{-.2cm}- B_{min},
\hspace{.1cm} j=1,2
\end{equation}
\begin{equation}\label{E9b}
\Delta B^z =  (B_{1x}^2+B_{1y}^2)^{1\over 2}-B_{min}
\end{equation}
\endnumparts
where $x_1=x$ and $x_2=y$.  Furthermore, the curvatures of the
magnetic field at the centre of the traps  can be written as
\numparts
\begin{equation}\label{E10a}
\frac{\partial^2 B}{\partial {x_j}^2}= \frac{4
\pi^2}{a^2}\frac{B_{0{x_j}}B_{1{x_j}}(B_{0x}B_{1x}
+B_{0y}B_{1y})}{({B_{0x}^2+B_{0y}^2})^{1\over 2}|B_{0x}B_{1y} -
B_{0y}B_{1x}|}, \quad j=1,2
\end{equation}
\begin{equation}\label{E10b}
\frac{\partial^2 B}{\partial z^2}= \frac{\partial^2 B}{\partial x^2}+\frac{\partial^2 B}{\partial y^2}
\end{equation}
\endnumparts
The trap frequencies, for an atom in hyperfine state $F$ with
magnetic quantum number $m_F$,  are given by
 \numparts
\begin{equation}\label{E11a}
\omega_{x_j}=  \frac{2\pi  \gamma}{a}
\left[\frac{B_{0x_j}B_{1x_j}(B_{0x}B_{1x}
+B_{0y}B_{1y})}{({B_{0x}^2+B_{0y}^2})^{1\over 2}|B_{0x}B_{1y} -
B_{0y}B_{1x}|}\right]^{1\over 2}, \quad j=1,2
\end{equation}
\begin{equation}\label{E11b}
\omega_z= \sqrt{ {\omega_x}^2+{\omega_y}^2}
\end{equation}
\endnumparts
where $\gamma$ = $\sqrt{{m_{{}_F}g_{{}_F}{\mu_{{}_B}}}/{m}}$,
$g_{{}_F}$ is the Land\'{e} g-factor, $\mu_{{}_B}$ is the Bohr
magneton and $m$ is the atomic mass.
\subsubsection{Symmetrical 2D magnetic lattice}\label{sec A symmetrical 2D}
To create a 2D lattice that is symmetrical with respect to $x$ and
$y$ we impose the constraint $\Delta B^x=\Delta B^y$ from which we
obtain $B_{1y} = \alpha_{0}B_{1x}$, where $\alpha_0=
{B_{0x}}/{B_{0y}}$, $B_{0x}\neq 0$ and $B_{0y}\neq 0$.  This is the
same as the condition given in~\cite{Ghanbari} for two crossed
layers of infinite periodic arrays of magnets  with bias fields
(\fref{figure2} (a) of~\cite{Ghanbari});  so our results are the
same as the equations given in~\cite{Ghanbari}. Here again, our
analytical expressions for $B_{min}$, $z_{min}$, $\Delta B^x$,
$\Delta B^y$, $\Delta B^z$, $\frac{\partial^2 B}{\partial x^2}$,
$\frac{\partial^2 B}{\partial y^2}$, $\frac{\partial^2 B}{\partial
z^2}$, $\omega_x$, $\omega_y$ and $\omega_z$ depend on the
$x$-component of the bias field $B_{1x}$ only, rather than on both
$B_{1x}$ and $B_{1y}$: \numparts
\begin{equation}\label{E12b}
B_{min}=\alpha_1 |B_{1x}|
\end{equation}
\begin{equation}\label{E12c}
z_{min}={a \over {2\pi}} \ln \left({{\alpha_2 B_{0x}}\over
|B_{1x}|}\right)
\end{equation}
\begin{equation}\label{E12d}
\frac{\partial^2 B}{\partial x^2}=\frac{\partial^2 B}{\partial
y^2}={1\over 2}\frac{\partial^2 B}{\partial z^2} = \frac{4 \pi^2
\alpha_3}{a^2} |B_{1x}|
\end{equation}
\begin{equation}\label{E12e}
\omega_x= \omega_y={\omega_z \over \sqrt{2}} = \frac{2\pi \gamma}{a}
\sqrt{\alpha_3 B_{1x}}
\end{equation}
\begin{equation}\label{E12f}
\Delta B^x = \Delta B^y =\alpha_4 |B_{1x}|
\end{equation}
\begin{equation}\label{E12g}
\Delta B^z=\alpha_5|B_{1x}|
\end{equation}
\endnumparts
where $B_{0x}$ and  $B_{0y}$ are given by \eref{E4a} and \eref{E4b}
and $\alpha_1$, $\cdots$, $\alpha_5$  are dimensionless constants
which depend on $\alpha_0$  and are given by \numparts
\begin{equation}\label{E13a}
\alpha_1= \frac{|1 - \alpha_0^2|}{{(1 +\alpha_0^2)}^{1\over 2}}
\end{equation}
\begin{equation}\label{E13b}
\alpha_2 = {1\over 2}\left(1 + \frac{1}{\alpha_0^2}\right)
\end{equation}
\begin{equation}\label{E13c}
\alpha_3=\frac{2
\alpha_0^2}{(1+\alpha_{0}^2)^{\frac{1}{2}}|1-\alpha_{0}^2|}
\end{equation}
\begin{equation}\label{E13d}
\alpha_4= \left(1 + \alpha_0^2 + \frac{4 \alpha_0^2 }{1 +
\alpha_0^2}\right)^{1\over 2} - \frac{|1 - \alpha_0^2|}{(1 +
\alpha_0^2)^{1\over 2}}
\end{equation}
\begin{equation}\label{E13e}
\alpha_5=(1+\alpha_{0}^2)^{1\over 2}-\frac{|1 - \alpha_0^2|}{{(1
+\alpha_0^2)}^{1\over 2}}
\end{equation}
\endnumparts
For a symmetrical 2D magnetic lattice with non-zero potential minima
above the surface of the top array, we have the constraints which
are the same as
 in~\cite{Ghanbari} \numparts
\begin{equation}\label{E14a}
\alpha_2 B_{0x}> |B_{1x}| > 0 , \qquad  \alpha_0 \alpha_2 B_{0y} >
|B_{1y}| > 0
\end{equation}
\begin{equation}\label{E14b}
B_{0x} \ne B_{0y}
\end{equation}
\endnumparts
\subsection{Special cases of magnetic lattices}\label{sec2.3}
In this section, we consider special cases of the  magnetic lattices
introduced in \sref{sec Four array of square}.
\subsubsection{ Single infinite periodic array of rectangular magnets}\label{sec2.3.1}
When $s_3=s_1=0$, $t_3=t_1=t$ and $t_4=t_2=0$ [\fref{figure1} (d)],
we obtain an infinite periodic array of rectangular magnets and
\eref{E4a}-\eref{E4c} give
\begin{equation}\label{E15}
B_{0x}=0,\qquad  B_{0y}=B_0({\rm e}^{kt}-1)
\end{equation}
\begin{figure}[tbp]
\begin{center}
$\begin{array}{cc}
\includegraphics[angle=0,width=6cm]{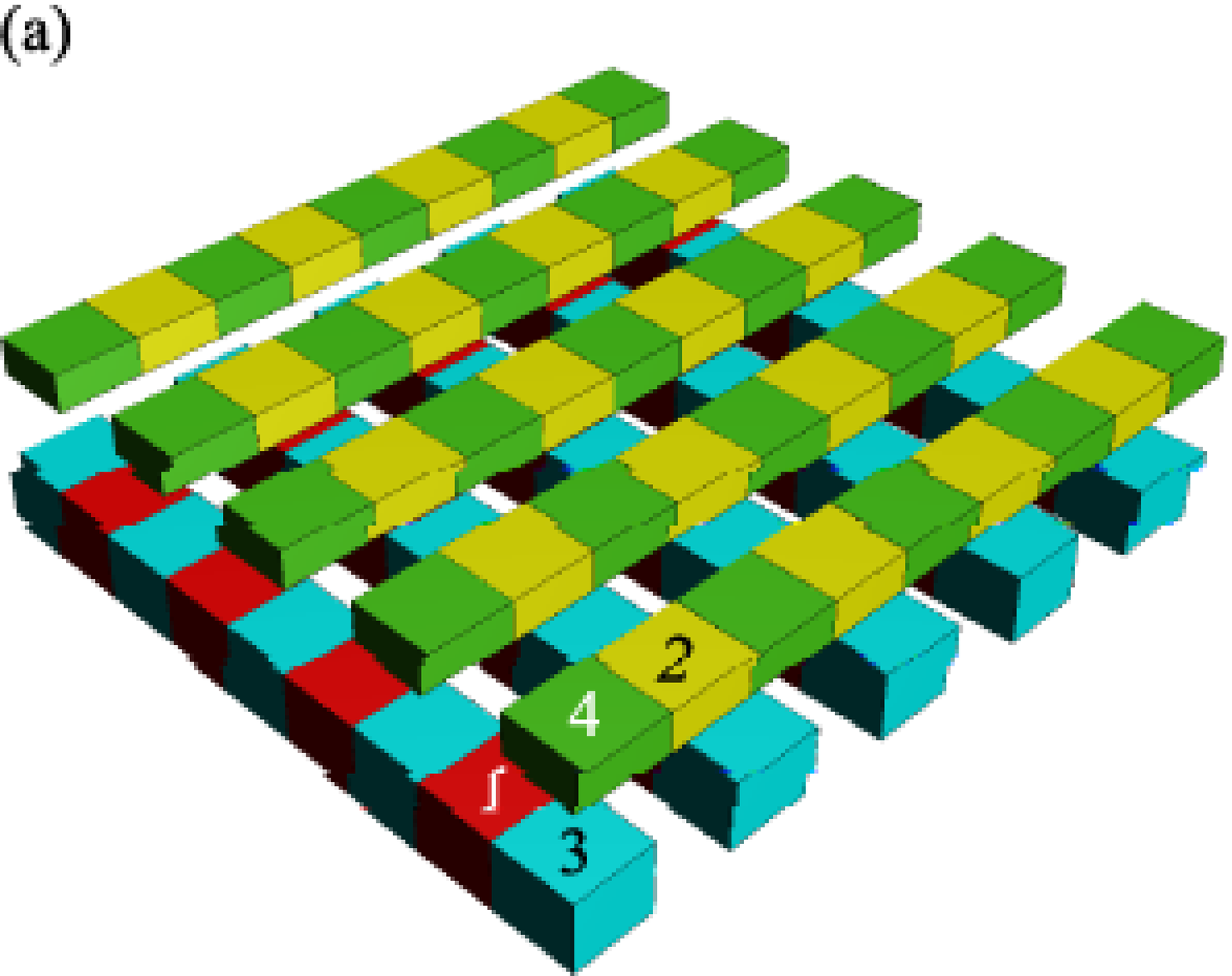}&
\includegraphics[angle=0,width=6cm]{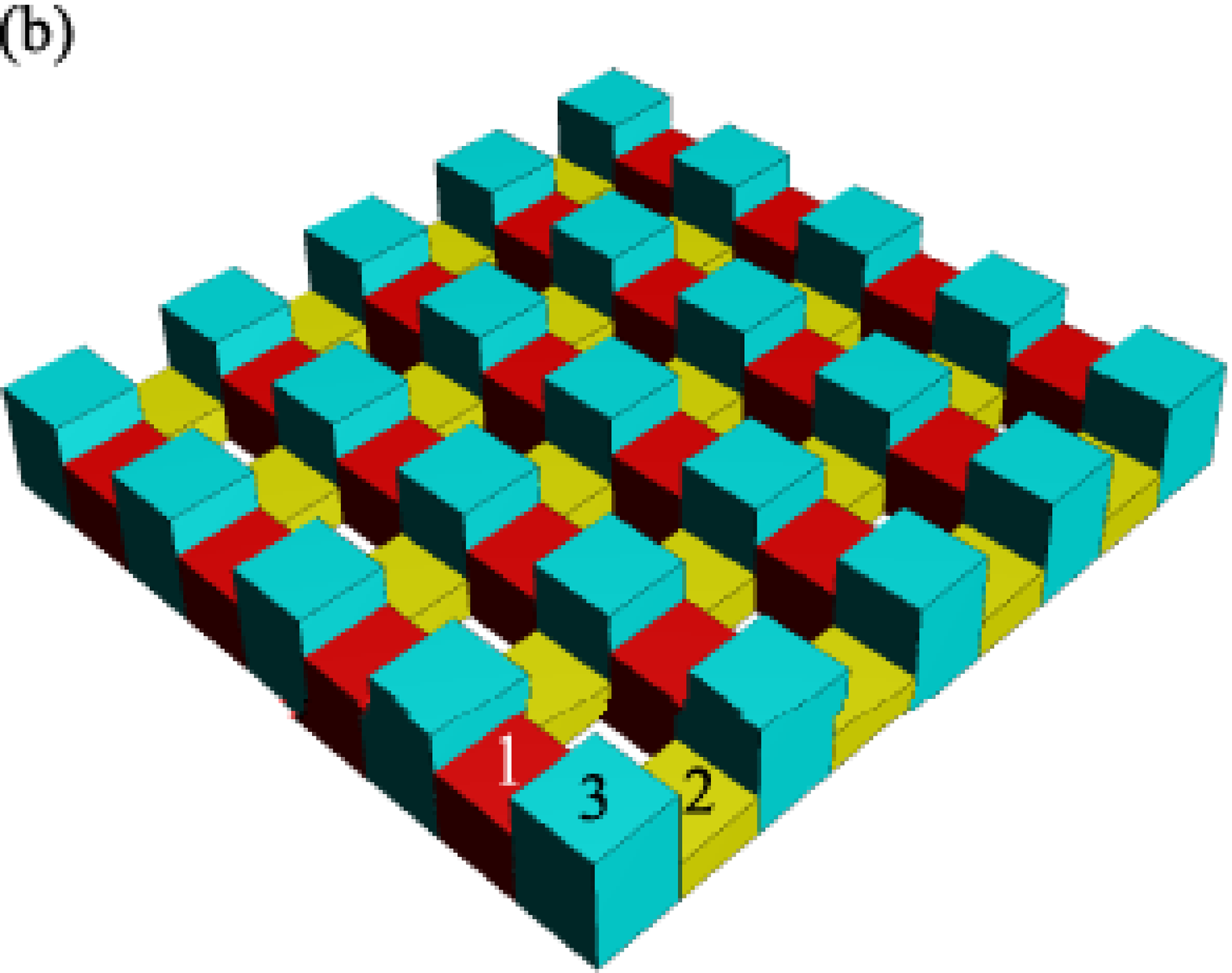}
\end{array}$
\caption{(a) Two crossed arrays of parallel  rectangular magnets
with perpendicular magnetization. (b) Three arrays  of square
magnets with different thickness.}\label{figure2}
\end{center}
\end{figure}
Substituting $B_{0x}$ and $B_{0y}$ from \eref{E15} into
\eref{E5}-\eref{E11b}, we obtain expressions for $B_{min}$,
$z_{min}$, $\Delta B^x$, $\Delta B^y$, $\Delta B^z$,
$\frac{\partial^2 B}{\partial x^2}$, $\frac{\partial^2 B}{\partial
y^2}$, $\frac{\partial^2 B}{\partial z^2}$, $\omega_x$, $\omega_y$
and $\omega_z$ which are the same as in~\cite{Ghanbari}.
\subsubsection{Two crossed arrays of parallel rectangular magnets}\label{sec Two crossed layers} Here, we
consider a configuration of square magnets in which $t_3 = t_1$,
$t_4 = t_2$, $s_2 =s_1=0$ and $s_4 = s_2 = s+ t_1$
[\fref{figure2}(a)].  From \eref{E4a}-\eref{E4c}, we have
\begin{equation}\label{E16}
B_{0x}= B_0({\rm e}^{kt_2}-1){\rm e}^{k(t_1+s)},\qquad
B_{0y}=B_0({\rm e}^{kt_1}-1)
\end{equation}
\Eref{E16} is in agreement with the expressions given
in~\cite{Ghanbari} where this magnetic lattice has been studied in
detail.
\begin{table}[hb]
 \caption{\label{table1}Numerical~\cite{Ghanbari} and analytical input parameters for a configuration
 consisting of three arrays of
 square magnets with different thickness
[\fref{figure2} (b)]}
\begin{indented}
\item[]
\hspace{-1.2cm}
\begin{tabular}{@{}llll}
\br
Parameter & Definition &  Numerical inputs &  Analytical inputs \\
\mr
\verb    $n_{sq}$    & Number of   squares                                     &    $401$   &   $\infty $              \\
\verb                                & in $x$- or $y$-direction                  &                      \\

\verb   $a\hspace{.1cm}(\mu m)$           & Period of magnetic lattice         &   $1.000 $   &   $1.000 $                   \\
\verb   $l_x=l_y\hspace{.1cm}(\mu m)$            &  Length of the chip       &   $200.5 $   &    $\infty $            \\
\verb                        &  along $x$ or $y$&                                 &     \\
\verb   $t_1\hspace{.1cm}(\mu m)$                &  Thickness of magnetic      &   $0.120 $      &     $0.120 $         \\
\verb                        &  film (first)                  &                                   &        \\
\verb   $t_2\hspace{.1cm}(\mu m)$                & Thickness of magnetic       &   $0.100 $        &    $0.100 $       \\
\verb                        &  film (second)                 &                                     &       \\
\verb   $t_3\hspace{.1cm}(\mu m)$                &  Thickness of  magnetic     &   $0.220 $          &    $0.220 $        \\
\verb                        & film (third)                   &                                   &                          \\
\verb   $4 \pi M_z\hspace{.1cm}(G)$          & Magnetization along z                       &   $3800 $  &     $3800 $                \\
\verb   $B_{1x}\hspace{.1cm}(G)$             & Bias magnetic field along $x$  &   $\hspace{-.33cm}$  $-5.00 $ &   $\hspace{-.33cm}$  $-5.00 $  \\
\verb   $B_{1y}\hspace{.1cm}(G)$             & Bias magnetic field along $y$   &   $\hspace{-.33cm}$ $-4.22 $  & $\hspace{-.33cm}$ $-4.22 $    \\
\verb   $B_{1z}\hspace{.1cm}(G)$             & Bias magnetic field along $z$   &   $\hspace{-.33cm}$ $-1.87 $   &  $0$ \\
\br
\end{tabular}
\end{indented}
\end{table}
\begin{table}[hb]
 \caption{\label{table2} Numerical~\cite{Ghanbari} and  analytical results
for the input parameters from \tref{table1} for a configuration
 consisting of three arrays  of
square magnets with different thickness [\fref{figure2} (b)]}
\begin{indented}
\item[]
\hspace{-2.4cm}
\begin{tabular}{@{}llll}
\br
Parameter & Definition &  Numerical results &  Analytical results \\
\mr
\verb   $x_{min}\hspace{.1cm}(\mu m)$       &  $x$ co-ordinate  of  potential  minimum       &    $0.250  $   &   $0.250  $      \\
\verb   $y_{min}\hspace{.1cm}(\mu m)$       &  $y$ co-ordinate  of   potential  minimum   &    $0.250  $   &   $0.250 $     \\
\verb   $z_{min}\hspace{.1cm}(\mu m)$        &  $z$ co-ordinate   of  potential   minimum   &     $0.952  $ &   $0.952  $  \\
\verb   $d\hspace{.1cm}(\mu m)$      &  Distance of  potential     &    $0.732$ &    $0.732$     \\
\verb            & minimum   from  surface  &    $$   &   $$     \\
\verb   $B_{min}\hspace{.1cm}(G)$       &  Magnetic field at   potential minimum  &    $1.10 $ &   $1.10$    \\
\verb   $\frac{\partial^2 B}{\partial x^2}\hspace{.1cm}({G\over{cm}^2})$       &   Curvature of $B$  along $x$   &     $7.52 \times 10^{10} $&   $7.52 \times 10^{10} $   \\
\verb   $\frac{\partial^2 B}{\partial y^2}\hspace{.1cm}({G\over{cm}^2})$       &   Curvature of $B$   along $y$  &     $7.52 \times 10^{10} $ &  $7.52 \times 10^{10} $   \\
\verb   $\frac{\partial^2 B}{\partial z^2}\hspace{.1cm}({G\over{cm}^2})$       &   Curvature of $B$ along $z$    &     $1.50 \times 10^{11} $  &   $1.50 \times 10^{11} $ \\
\verb   $\Delta B^x\hspace{.1cm}(G)$       &  Magnetic barrier   height along $x$     &       $8.10$ &   $8.10$  \\
\verb   $ \Delta B^y\hspace{.1cm}(G)$       &  Magnetic barrier   height along $y$     &     $8.09$ &   $8.10$    \\
\verb   $\Delta B^z\hspace{.1cm}(G)$       &  Magnetic barrier    height along $z$  &     $5.45$ &   $5.45$   \\
\verb   $U_{min}/k_B \hspace{0.1cm}(\mu { K})$    & Potential energy minimum   &   $74  $  &     $73$          \\
\verb   $\Delta U^x/k_B \hspace{0.1cm}(\mu { K})$    & Potential barrier height along $x$    &   $544 $   &     $544 $          \\
\verb   $\Delta U^y/k_B \hspace{0.1cm}(\mu { K})$    & Potential barrier height along $y$    &   $544 $    &     $544 $         \\
\verb   $\Delta U^z/k_B \hspace{0.1cm}(\mu { K})$    & Potential barrier height along $z$    &   $366 $     &    $366 $        \\
\verb   $\omega_x/ 2\pi \hspace{0.1cm}({ kH}z)$   &  Trap frequency along $x$            &    $350 $    &    $350 $        \\
\verb   $\omega_y/ 2\pi \hspace{0.1cm}({ kH}z)$   &  Trap frequency along $y$            &    $350 $     &     $350 $      \\
\verb   $\omega_z/ 2\pi \hspace{0.1cm}({ kH}z)$   &  Trap frequency along $z$              &   $494 $     &     $495$        \\
\verb   $\hbar\omega_x/k_B \hspace{0.1cm}(\mu { K})$ &  Level spacing along $x$               &   $17 $    &      $17 $         \\
\verb   $\hbar\omega_y/k_B \hspace{0.1cm}(\mu { K})$ &  Level spacing along $y$               &   $17 $     &      $17 $        \\
\verb   $\hbar\omega_z/k_B \hspace{0.1cm}(\mu { K})$ &  Level spacing along $z$                &   $24 $     &     $24 $         \\
\br
\end{tabular}
\end{indented}
\end{table}
\subsubsection{Three arrays of square magnets with different thickness}\label{sec2.2.3} When $s_1 = s_2= s_3=t_4 =0$
[\fref{figure2}(b)], from \eref{E4a}-\eref{E4c}, we have
\begin{equation}\label{E17}
B_{0x}= {B_0\over 2}(-{\rm e}^{kt_1}+{\rm e}^{kt_2}+{\rm
e}^{kt_3}-1),\quad B_{0y}={B_0\over 2}({\rm e}^{kt_1}-{\rm
e}^{kt_2}+{\rm e}^{kt_3}-1)
\end{equation}
We note that, according to \eref{E14b}, for a symmetrical 2D lattice
with  non-zero magnetic potential minima, we should have $t_2 \neq
t_1$. Using equations \eref{E5}-\eref{E11b} and \eref{E17} we can
determine analytically  the relevant quantities for this system.
This magnetic lattice has already been studied
numerically~\cite{Ghanbari}. \Tref{table1} gives the
numerical~\cite{Ghanbari} and analytical inputs and \tref{table2}
gives calculated values for the relevant quantities. There is good
agreement between  the analytical results for an infinite lattice
and the numerical results~\cite{Ghanbari} which were obtained in the
central region of the lattice.
\subsubsection{Chessboard configuration of square magnets}\label{sec2.2.5}
In \fref{figure2}(b), if we take $t_1 = t_2$ and $ t_3=0$, we have a
chessboard configuration of square magnetic slabs. In this case,
from \eref{E17}, we obtain
\begin{equation}\label{E18}
B_{0x}= B_{0y}=0
\end{equation}
Considering \eref{E14b}, we do not have  a lattice of {\it non-zero}
magnetic field minima. In fact, we do not have  a magnetic lattice,
because according to \eref{E3a}-\eref{E3c} we obtain a uniform
magnetic field if we look at  points  sufficiently far  from the
surface of the array.
\begin{figure}[tbp]
\begin{center}
$\begin{array}{c}
\includegraphics[angle=0,width=7cm]{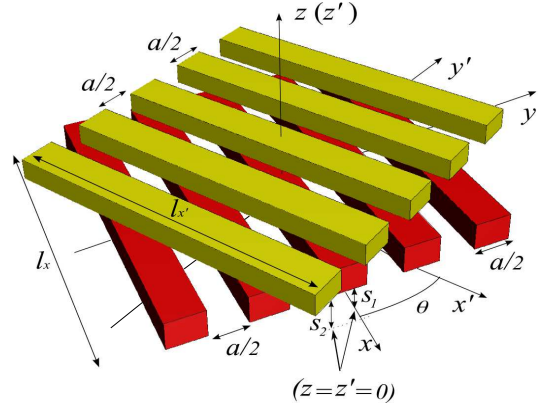}
\end{array}$
\caption{Two crossed arrays of parallel rectangular magnets at an
arbitrary angle $\theta$ with respect to each other.}\label{figure3}
\end{center}
\end{figure}
\section{Two crossed periodic arrays of parallel rectangular magnets at an arbitrary angle with respect to each other}
\subsection{General case}\label{General lattice}
Here, we consider a configuration of magnets which is a
generalization of \sref{sec Two crossed layers}. \Fref{figure3}
shows this configuration of rectangular magnetic slabs in which the
upper array of parallel magnets is rotated at an arbitrary angle
$\theta$ about the $z$ axis which is normal to the  surface. The
magnets in the lower array are parallel to the $x$-axis while the
magnets in the upper array are parallel to the $x^{\prime}$-axis. If
$\theta=\pi /2$, according to \fref{figure3}, we have the two
crossed infinite periodic arrays of magnets shown in \fref{figure2}
(a). Considering the symmetry, in the rotated coordinate system of
reference, we have for the upper array \numparts
\begin{equation}\label{E19a}
B_{{x}^{\prime}} =  0
\end{equation}
\begin{equation}\label{E19b}
B_{{y}^{\prime}}= B_{0{{y}^{\prime}}} \sin (k{{y}^{\prime}}) {\rm
e}^{-k{{z}^{\prime}}}
\end{equation}
\begin{equation}\label{E19c}
B_{{z}^{\prime}} =  B_{0{{y}^{\prime}}}  \cos (k{{y}^{\prime}}) {\rm
e}^{-k{{z}^{\prime}}}
\end{equation}
\endnumparts
where $ B_{0{y}^{\prime}} = B_{0} ( {\rm e}^{kt_2}-1){\rm
e}^{ks_2}$,  ${{y}^{\prime}}=-x \sin \theta + y \cos \theta$ and
$z^{\prime}=z$. Now, we can write the components of the magnetic
field in the original coordinate system \numparts
\begin{equation}\label{E20a}
B_{{x}} =  B_{0{{y}^{\prime}}} \sin\theta \sin (k{{y}^{\prime}})
{\rm e}^{-k{{z}}}
\end{equation}
\begin{equation}\label{E20b}
B_{{y}}= B_{0{{y}^{\prime}}}\cos\theta \sin (k{{y}^{\prime}}) {\rm
e}^{-k{{z}}}
\end{equation}
\begin{equation}\label{E20c}
B_{{z}} = B_{0{{y}^{\prime}}}  \cos (k{{y}^{\prime}}) {\rm
e}^{-k{{z}}}
\end{equation}
\endnumparts
The components of the total magnetic field due to both  arrays of
parallel magnets plus a bias field ${\bf B_1}=B_{1x}{\hat
x}+B_{1y}{\hat y}$ are then \numparts
\begin{equation}\label{E21a}
B_{{x}} =  - B_{0{{y}^{\prime}}} \sin\theta \sin (k{{y}^{\prime}})
{\rm e}^{-k{{z}}} + B_{{1x}}
\end{equation}
\begin{equation}\label{E21b}
B_{{y}}= [B_{0{{y}^{\prime}}}\cos\theta \sin (k{{y}^{\prime}})+
B_{0{{y}}} \sin (k{{y}})] {\rm e}^{-k{{z}}}+B_{{1y}}
\end{equation}
\begin{equation}\label{E21c}
B_{{z}} = [B_{0{{y}^{\prime}}}  \cos (k{{y}^{\prime}}) +B_{0{{y}}}
\cos (k{{y}}) ]{\rm e}^{-k{{z}}}
\end{equation}
\endnumparts
where $z \gg {a\over{2\pi}}+ s_2 + t_2$ and $B_{0{{y}}}= B_{0} (
{\rm e}^{kt_1}-1){\rm e}^{ks_1}$.  The magnitude of the magnetic
field is then
\begin{eqnarray}
 B(x,y,z) &=& \Bigl\{B_{1x}^2+B_{1y}^2 +  2 [B_{0y^{\prime}}(-B_{1x} \sin \theta  \nonumber\\ 
  &+& B_{1y} \cos \theta ) \sin(ky^{\prime})+ B_{0y}B_{1y} \sin(ky)]{\rm
 e}^{-kz}\nonumber\\
  &+& (B_{0y^{\prime}}^2+B_{0y}^2 + 2 B_{0y^{\prime}}B_{0y}
[\cos(ky^{\prime})\cos(ky)\nonumber\\
 &+& \sin(ky^{\prime})\sin(ky)\cos\theta]){\rm e}^{-2kz}
\Bigr\}^{1\over 2}\label{E22}
\end{eqnarray}
If $\theta > 0$, these arrays of rectangular magnets can give a 2D
periodic lattice of magnetic traps with {\it non-zero} potential
minima given by
\begin{equation}\label{E23}
 B_{min}=  \frac{|-B_{0{{y}}}B_{1{{x}}}+ B_{0{y}^{\prime}}B_{1x}\cos \theta+B_{0{y}^{\prime}}B_{1y}\sin^2 \theta|
}{({B_{0{{y}^{\prime}}}^2+B_{0y}^2-2
B_{0{{y}^{\prime}}}B_{0{{y}}}\cos \theta})^{1\over 2}}
\end{equation}
The potential minima are located at \numparts
\begin{equation}\label{E24a}
x_{min}=\left(n_x+{1 \over 4}\right) {{a \sin \theta}\over {1-\cos
\theta}}, \qquad n_x=0,\pm 1,\pm 2,\cdots
\end{equation}
\begin{equation}\label{E24b}
y_{min}=\left(n_y+{1 \over 4}\right) a, \hspace{2.4cm}  n_y=0,\pm
1,\pm 2,\cdots
\end{equation}
\begin{equation}\label{E24c}
z_{min}={a \over {2\pi}} \ln \left({{ B_{0{{y}^{\prime}}}^2+B_{0y}^2
-2 B_{0{{y}^{\prime}}}B_{0{{y}}}\cos \theta}\over
-B_{0{{y}^{\prime}}}B_{1{{x}}}\sin\theta+B_{0y^{\prime}}B_{1y}\cos\theta-B_{0y}B_{1y}}\right)
\end{equation}
\begin{equation}\label{E24d}
x_{min}^{\prime}=\left(n_{x^{\prime}}+{1 \over 4}\right) {{a \sin
\theta}\over {1-\cos \theta}}, \qquad n_{x^{\prime}}=0,\pm 1,\pm
2,\cdots
\end{equation}
\begin{equation}\label{E24e}
y_{min}^{\prime}=\left(n_{y^{\prime}}-{1 \over 4}\right) a,
\hspace{2.4cm} n_{y^{\prime}}=0,\pm 1,\pm 2,\cdots
\end{equation}
\endnumparts
The period $p$ of the magnetic potential minima, which are located
along  straight lines parallel to the $x$- and $x^{\prime}$-axes
(defined in \fref{figure3}), varies with $\theta$ and is given by
\begin{equation}\label{E25}
p=\left|{{a \sin \theta}\over {1-\cos \theta}}\right|
\end{equation}
If $\theta$ changes between $0$ and $\pi/2$, the period of the
magnetic lattice varies between $\infty$ and the minimum period $a$,
which is the period of the magnetic slabs along the $y$- and
$y^{\prime}$-axes.
\begin{figure}[tbp]
\begin{center}
$\begin{array}{c}
\includegraphics[angle=0,width=12.9cm]{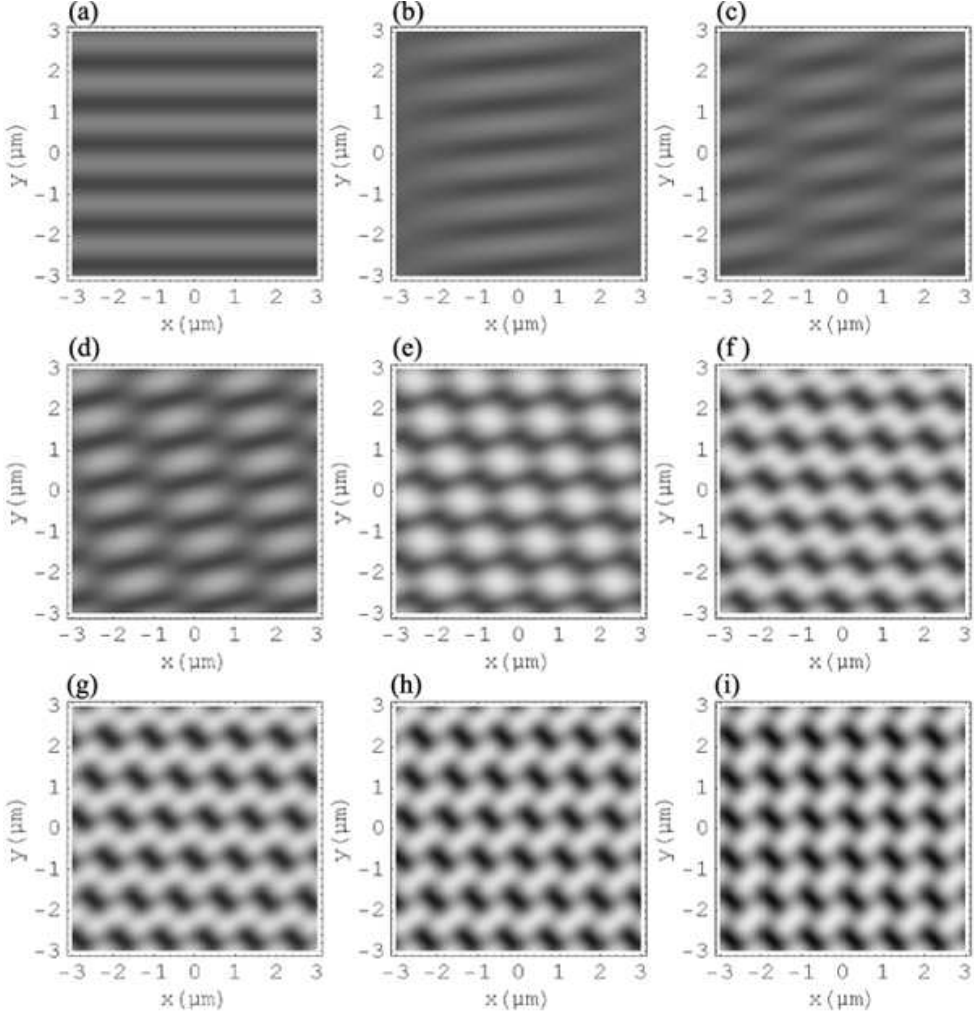}
\end{array}$
\caption{Density plot of the  magnetic field, calculated using
analytical expressions, for two crossed arrays  of parallel
rectangular magnets at an arbitrary angle $\theta$ with respect to
each other for the input parameters from \tref{table3} and (a)
$\theta=0$ (b) $\theta=\pi/18$ (c) $\theta=\pi/9$
 (d) $\theta=\pi/6$ (e) $\theta=\pi/4$ (f) $\theta=\pi/3$
 (g) $\theta=7\pi/18$ (h) $\theta=4\pi/9$ (i)
 $\theta=\pi/2$}\label{figure4}.
\end{center}
\end{figure}
Now, we can write the constraints for a lattice with  {\it non-zero}
magnetic field minima
 \numparts
\begin{equation}\label{E26a}
-B_{0{{y}}}B_{1{{x}}}+ B_{0{y}^{\prime}}B_{1x}\cos
\theta+B_{0{y}^{\prime}}B_{1y}\sin^2 \theta\neq 0
\end{equation}
\begin{equation}\label{E26b}
\hspace{-2.0cm}{ B_{0{{y}^{\prime}}}^2+B_{0y}^2 -2
B_{0{{y}^{\prime}}}B_{0{{y}}}\cos
\theta}>-B_{0{{y}^{\prime}}}B_{1{{x}}}\sin\theta+B_{0y^{\prime}}B_{1y}\cos\theta-B_{0y}B_{1y}
>0
\end{equation}
\endnumparts
According to \eref{E23}-\eref{E24c} and the contour plots of the
magnetic field for different values of $\theta$ (\fref{figure4}),
apart from $\theta=0$ for which we have a 1D array of 2D magnetic
microtraps, we have a 2D array of 3D microtraps with different
periods in the $x$- and $x^{\prime}$-directions, as $\theta$ varies.
Looking at \fref{figure4} more carefully, we find that as $\theta$
changes the distance between the magnetic traps along the  interior
bisector of the angle $\theta$ between the positive $y$- and
$y^{\prime}$-axes also changes.  Thus, depending on the barrier
heights, the tunnelling rate between the traps
 in this direction will be higher than the  tunnelling rate in the $x$- and
 $x^{\prime}$-directions. Here, we give the barrier heights in the $x$-,
$x^{\prime}$- and $z$-directions \numparts
\begin{equation*}
  \hspace{-2cm}\Delta B^{x_j} = \Bigl\{\left[ B_{1x}-(-1)^j {{B_{0y^{\prime}}\sin\theta (B_{0y}
 B_{1y}-B_{0y^{\prime}}B_{1y}\cos\theta+B_{0y^{\prime}}B_{1x}\sin\theta)}\over {{B_{0{{y}^{\prime}}}^2+B_{0y}^2-2
B_{0{{y}^{\prime}}}B_{0{{y}}}\cos \theta}}} \right]^2  \nonumber
\end{equation*}
\begin{equation*}
\hspace{-2cm}+ \left[ B_{1y}+(-1)^j {{(B_{0y}+
B_{0y^{\prime}}\cos\theta) (B_{0y}
 B_{1y}-B_{0y^{\prime}}B_{1y}\cos\theta+B_{0y^{\prime}}B_{1x}\sin\theta)}\over {{B_{0{{y}^{\prime}}}^2+B_{0y}^2-2
B_{0{{y}^{\prime}}}B_{0{{y}}}\cos \theta}}} \right]^2
\Bigr\}^{1\over 2}
\end{equation*}
\begin{equation}\label{E27a}
\hspace{0cm}-B_{min}, \qquad \qquad \qquad \qquad \qquad  j=1
\hspace{.2cm} {\rm or}\hspace{.2cm} 4
\end{equation}
\begin{equation}\label{E27b}
\Delta B^z =  (B_{1x}^2+B_{1y}^2)^{1\over 2}-B_{min}
\end{equation}
\endnumparts
where, $x_1=x$, $x_4=x^{\prime}$ and by definition we have
 \numparts
\begin{equation}\label{E28a}
\Delta B^{x} = B(x=x_{max},y=y_{min},z_{min}) -B_{min}
\end{equation}
\begin{equation}\label{E28b}
\Delta B^{x^{\prime}} =
B(x=x_{max}^{\prime},y=y_{min}^{\prime},z_{min}) -B_{min}
\end{equation}
\begin{equation}\label{E28c}
\Delta B^{z} =  B(x_{min},y_{min},z=\infty)-B_{min}
\end{equation}
\endnumparts
where
\numparts
\begin{equation}\label{E29a}
x_{max}={{a \sin \theta}\over {1-\cos \theta}}\left(n_{x}+{3 \over
4}\right) , \qquad n_{x}=0,\pm 1,\pm 2,\cdots
\end{equation}
\begin{equation}\label{E29b}
x_{max}^{\prime}={{a \sin \theta}\over {1-\cos
\theta}}\left(n_{x^{\prime}}+{3 \over 4}\right) , \qquad
n_{x^{\prime}}=0,\pm 1,\pm 2,\cdots
\end{equation}
\endnumparts
We also have \numparts
\begin{equation}\label{E30a}
\Delta B^{x} = B(y^{\prime}={a\over 4},y={a\over 4},z_{min})
-B_{min}
\end{equation}
\begin{equation}\label{E30b}
\Delta B^{x^{\prime}} = B(y^{\prime}=-{{3 a}\over 4},y={a\over
4},z_{min})-B_{min}
\end{equation}
\endnumparts
Furthermore,  the curvatures are  given by \numparts
\begin{equation*}
\hspace{-.9cm}\frac{\partial^2 B}{\partial x^2}=\frac{
\pi^2}{a^2{F^{3\over
2}G}}B_{0y^{\prime}}\sin^2\theta[2B_{0y^{\prime}}(B_{0y^{\prime}}^2+B_{0y}^2)(B_{1x}^2+B_{1y}^2)
\end{equation*}
\begin{equation*}
-B_{0y}B_{0y^{\prime}}^2(B_{1x}^2+7B_{1y}^2)\cos\theta-2B_{0y^{\prime}}(B_{0y^{\prime}}^2+B_{0y}^2)(B_{1x}^2-B_{1y}^2)\cos
2 \theta
\end{equation*}
\begin{equation*}
+B_{0y}B_{0y^{\prime}}^2(B_{1x}^2-B_{1y}^2)\cos 3 \theta + 2
B_{0y}(3B_{0y^{\prime}}^2+2B_{0y}^2)B_{1x}B_{1y}\sin\theta
\end{equation*}
\begin{equation}\label{E31a}
-4B_{0y^{\prime}}(B_{0y^{\prime}}^2+B_{0y}^2)B_{1x}B_{1y}\sin
2\theta+2B_{0y}B_{0y^{\prime}}^2B_{1x}B_{1y}\sin 3\theta]
\end{equation}
\begin{equation*}
\hspace{-.9cm}\frac{\partial^2 B}{\partial {x^{\prime}}^2}= \frac{
\pi^2}{a^2{F^{3\over 2}G}}B_{0y}\sin^2\theta [ B_{0y^{\prime}}(-
8{B_{0y}}^2 {B_{1y}}^2 + B_{0y^{\prime}}^2
(B_{1x}^2-B_{1y}^2)\cos\theta
\end{equation*}
\begin{equation*}
-B_{0y^{\prime}}^3 (B_{1x}^2-B_{1y}^2)\cos 3\theta +
4B_{1y}(B_{0y}(B_{0y}^2+{B_{0y^{\prime}}}^2)B_{1y}
\end{equation*}
\begin{equation}\label{E31b}
+B_{0y^{\prime}}B_{1x}(B_{0y}^2-{B_{0y^{\prime}}}^2 \cos 2\theta)
\sin \theta )]
\end{equation}
\begin{equation*}
\hspace{-.9cm}\frac{\partial^2 B}{\partial {y}^2}=\frac{2
\pi^2}{a^2{F^{3\over 2}G}}(B_{0y}B_{1y}
-B_{0y^{\prime}}B_{1y}\cos\theta+B_{0y^{\prime}}B_{1x}\sin\theta)[
\end{equation*}
\begin{equation*}
(B1y(2B0y^3+B_{0y^{\prime}}\cos\theta(-6B_{0y}^2+B_{0y^{\prime}}\cos\theta(-2B_{0y^{\prime}}\cos\theta
\end{equation*}
\begin{equation*}
+B_{0y}(5+\cos
2\theta)))))+2B_{0y^{\prime}}B_{1x}\cos\theta(B_{0y^{\prime}}-B_{0y}\cos\theta)
\end{equation*}
\begin{equation}\label{E31c}
(-B_{0y}+B_{0y^{\prime}}\cos\theta)\sin\theta
+2B_{0y}B_{0y^{\prime}}^2B_{1y}\sin^2\theta]
\end{equation}
\begin{equation*}
\hspace{-.9cm}\frac{\partial^2 B}{\partial {y^{\prime}}^2}=\frac{
\pi^2}{a^2{F^{3\over 2}G}}(B_{0y}B_{1y}
-B_{0y^{\prime}}B_{1y}\cos\theta+B_{0y^{\prime}}B_{1x}\sin\theta)[
\end{equation*}
\begin{equation*}
2B_{1y} \cos\theta ( -5B_{0y}^2B_{0y^{\prime}}
-2{B_{0y^{\prime}}}^3+2B_{0y}^3\cos\theta
+7B_{0y}{B_{0y^{\prime}}}^2 \cos\theta
\end{equation*}
\begin{equation*}
-B_{0y}B_{0y^{\prime}}(B_{0y}+B_{0y^{\prime}}\cos\theta)\cos
2\theta)+B_{0y^{\prime}}B_{1x}(4({B_{0y}}^2+{B_{0y^{\prime}}}^2)
\end{equation*}
\begin{equation}\label{E31d}
+B_{0y}B_{0y^{\prime}}(-9\cos\theta+\cos 3\theta))\sin\theta]
\end{equation}
\begin{equation}\label{E31e}
\hspace{-.9cm}\frac{\partial^2 B}{\partial z^2}= \frac{4
\pi^2}{a^2{F^{1\over 2}G}}(B_{0y}B_{1y}
-B_{0y^{\prime}}B_{1y}\cos\theta+B_{0y^{\prime}}B_{1x}\sin\theta)^2
\end{equation}
\endnumparts
where \numparts
\begin{equation}\label{E32a}
F=|{B_{0y^{\prime}}^2+B_{0y}^2- 2B_{0y}B_{0y^{\prime}}\cos\theta}|
\end{equation}
\begin{equation}\label{E32b}
G=|-B_{0y}B_{1x}+B_{0y^{\prime}}B_{1x}\cos\theta
+B_{0y^{\prime}}B_{1y}\sin\theta|
\end{equation}
\endnumparts
\begin{table}[hb]
 \caption{\label{table3}Numerical and analytical input parameters for  two crossed arrays of parallel rectangular magnets
 at an angle $\theta = {\pi/6}$ with respect to each other
(\fref{figure3}): (1) numerical inputs (2) analytical inputs}
\begin{indented}
\item[]
\hspace{-2.0cm}
\begin{tabular}{@{}llll}
\br
Parameter & Definition &  Numerical inputs &  Analytical inputs \\
\mr
\verb    $n_{r}$    & Number of   parallel rectangular                                    &    $1001$   &   $\infty $              \\
\verb                                &magnets   in $x$- or $x^{\prime}$-direction                  &                      \\

\verb   $a\hspace{.1cm}(\mu m)$           & Period of magnetic lattice         &   $1.000 $   &   $1.000 $                   \\
\verb   $l_x=l_x^{\prime}\hspace{.1cm}(\mu m)$            &  Length of the magnets       &   $1000.5 $   &    $\infty $            \\
\verb                        &  along $x$- or $x^{\prime}$-direction&                                 &     \\
\verb   $t_1\hspace{.1cm}(\mu m)$                &  Thickness of magnetic      &   $0.322 $      &     $0.322 $         \\
\verb                        &  slabs parallel to the $x$-direction                  &                                   &        \\
\verb   $t_2\hspace{.1cm}(\mu m)$                & Thickness of magnetic       &   $0.083 $        &    $0.083 $       \\
\verb                        &   slabs parallel to the $x^{\prime}$-direction                  &                                     &       \\

\verb   $s_1\hspace{.1cm}(\mu m)$                & Distance from the plane       &   $0.000 $        &    $0.000 $       \\
\verb                        &$z=0$    to the lower surface of                   &                                     &       \\
\verb                        & the   magnets parallel                    &                                     &       \\
\verb                        &   to the  $x$-direction                   &                                     &       \\
\verb   $s_2\hspace{.1cm}(\mu m)$                & Distance from the plane       &   $0.422 $        &    $0.422 $       \\
\verb                        &$z=0$    to the lower surface of                   &                                     &       \\
\verb                        & the   magnets parallel                    &                                     &       \\
\verb                        &   to the  $x^{\prime}$-direction                   &                                     &       \\
\verb   $4 \pi M_z\hspace{.1cm}(G)$          & Magnetization along z                       &   $3800 $  &     $3800 $                \\
\verb   $B_{1x}\hspace{.1cm}(G)$             & Bias magnetic field along $x$  &   $\hspace{-.33cm}$  $-4.08 $ &   $\hspace{-.33cm}$  $-4.08$  \\
\verb   $B_{1y}\hspace{.1cm}(G)$             & Bias magnetic field along $y$   &   $\hspace{-.33cm}$ $-1.32 $  & $\hspace{-.33cm}$ $-1.32 $    \\
\verb   $B_{1z}\hspace{.1cm}(G)$             & Bias magnetic field along $z$   &   $\hspace{-.33cm}$ $-0.69 $   &  $0.00$ \\
\br
\end{tabular}
\end{indented}
\end{table}
\begin{table}[hb]
 \caption{\label{table4} Numerical and  analytical results
for the input parameters from \tref{table3} for  two crossed arrays
of parallel rectangular magnets
 at an angle $\theta = {\pi/6}$ with respect to each other
(\fref{figure3})}
\begin{indented}
\item[]
\hspace{-2.4cm}
\begin{tabular}{@{}llll}
\br
Parameter & Definition &  Numerical results &  Analytical results \\
\mr
 $x_{min}\hspace{.1cm}(\mu m)$       &  $x$ co-ordinate  of  potential   minimum      &    $0.933  $   &   $0.933  $      \\
   $y_{min}\hspace{.1cm}(\mu m)$       &  $y$ co-ordinate  of  potential  minimum   &    $0.250  $   &   $0.250 $     \\
 $z_{min}\hspace{.1cm}(\mu m)$        &  $z$ co-ordinate  of  potential minimum     &     $1.200  $ &   $1.200  $  \\
   $d\hspace{.1cm}(\mu m)$      &  Distance of  potential    &    $0.695$ &    $0.695$     \\
            &  minimum  from  surface   &    $$   &   $$     \\
  $B_{min}\hspace{.1cm}(G)$       &  Magnetic field at   potential minimum    &    $2.68 $ &   $2.68$    \\
   $\frac{\partial^2 B}{\partial x^2}\hspace{.1cm}({G\over{cm}^2})$       &   Curvature of $B$  along $x$  &     $1.05 \times 10^{10} $&   $1.05 \times 10^{10} $   \\
   $\frac{\partial^2 B}{\partial y^2}\hspace{.1cm}({G\over{cm}^2})$       &   Curvature of $B$ along $y$  &     $5.99 \times 10^{9} $ &  $5.98 \times 10^{9} $   \\
   $\frac{\partial^2 B}{\partial z^2}\hspace{.1cm}({G\over{cm}^2})$       &   Curvature of $B$  along $z$     &     $1.65 \times 10^{10} $  &   $1.65 \times 10^{10} $ \\
   $\Delta B^x\hspace{.1cm}(G)$       &  Magnetic barrier  height along $x$     &       $8.32$ &   $8.33$  \\
   $\Delta B^{x^{\prime}}\hspace{.1cm}(G)$       &  Magnetic barrier    height along $x^{\prime}$   &     $8.31$ &   $8.33$    \\
   $\Delta B^z\hspace{.1cm}(G)$       &  Magnetic barrier    height along $z$   &     $1.60$ &   $1.60$   \\
   $U_{min}/k_B \hspace{0.1cm}(\mu { K})$    & Potential energy minimum           &   $180  $  &     $180$          \\
   $\Delta U^x/k_B \hspace{0.1cm}(\mu { K})$    & Potential barrier height along $x$    &   $559 $   &     $559 $          \\
   $\Delta U^{x^{\prime}}/k_B \hspace{0.1cm}(\mu { K})$    & Potential barrier height along $x^{\prime}$    &   $558 $    &     $559 $         \\
   $\Delta U^z/k_B \hspace{0.1cm}(\mu { K})$    & Potential barrier height along $z$    &   $108$     &    $108 $        \\
   $\omega_x/ 2\pi \hspace{0.1cm}({ kH}z)$   &  Trap frequency along $x$            &    $131 $    &    $131 $        \\
  $\omega_y/ 2\pi \hspace{0.1cm}({ kH}z)$   &  Trap frequency along $y$            &    $99 $     &     $99 $      \\
   $\omega_z/ 2\pi \hspace{0.1cm}({ kH}z)$   &  Trap frequency along $z$              &   $164 $     &     $164$        \\
   $\hbar\omega_x/k_B \hspace{0.1cm}(\mu { K})$ &  Level spacing along $x$               &   $6 $    &      $6 $         \\
   $\hbar\omega_y/k_B \hspace{0.1cm}(\mu { K})$ &  Level spacing along $y$               &   $5 $     &      $5 $        \\
   $\hbar\omega_z/k_B \hspace{0.1cm}(\mu { K})$ &  Level spacing along $z$                &   $8 $     &     $8 $         \\
\br
\end{tabular}
\end{indented}
\end{table}
\subsection{Symmetrical lattice}\label{Symmetrical lattice}
\begin{figure}[tbp]
\begin{center}
$\begin{array}{c}
\includegraphics[angle=0,width=13.3cm]{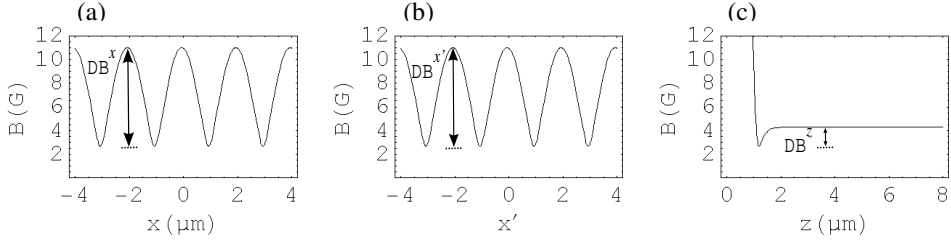}
\end{array}$
\caption{Magnetic field produced by  two crossed arrays of parallel
rectangular magnets at an angle $\theta = \pi/6$ with respect to
each other, with input parameters from \tref{table3}. Curves (a)-(c)
show the analytical results of the magnetic field  (a) along a line
($y = y_{min}$, $z = z_{min}$) parallel to the $x$-axis, (b) along a
line ($y^{\prime}=y^{\prime}_{min}$, $z=z_{min}$) parallel to the
$x^{\prime}$-axis, and (c) along a line ($x=x_{min}$, $y=y_{min}$)
parallel to the $z$-axis.}\label{figure5}
\end{center}
\end{figure}
For $\theta \neq 0$, if we have $B_{1y} = \beta_{0}(\theta)B_{1x}$,
where
$\beta_{0}(\theta)={{B_{0y^{\prime}}\sin\theta}/(B_{0y^{\prime}}\cos\theta+B_{0y}})$,
then the barrier heights in the $x$- and $x^{\prime}$-directions are
equal. Thus,  we have a symmetrical 2D lattice of 3D microtraps in
the $x$- and $x^{\prime}$-directions and, as \fref{figure5} shows,
the microtraps in the $x$- and $x^{\prime}$-directions have the same
period. This figure shows  our analytical results for an infinite
lattice, which  are in excellent agreement with our numerical
calculations in the central region of the lattice.
 For the numerical calculations we have
written a Mathematica code and have used  the software package
Radia~\cite{esrf} interfaced to Mathematica. Here, the quantities of
interest are \numparts
\begin{equation}\label{E33a}
B_{min}=\beta_1(\theta) |B_{1x}|
\end{equation}
\begin{equation}\label{E33b}
z_{min}={a \over {2\pi}} \ln \left[{{\beta_2(\theta) B_{0y}}\over
|B_{1x}|}\right]
\end{equation}
\begin{equation}\label{E33c}
\frac{\partial^2 B}{\partial x^2}=\frac{\partial^2 B}{\partial
{x^{\prime}}^2}= \frac{4 \pi^2 \beta_3(\theta)}{a^2} |B_{1x}|
\end{equation}
\begin{equation}\label{E33d}
\frac{\partial^2 B}{\partial y^2} = \frac{\partial^2 B}{\partial
{y^{\prime}}^2}=\frac{4 \pi^2 \beta_4(\theta)}{a^2} |B_{1x}|
\end{equation}
\begin{equation}\label{E33e}
\frac{\partial^2 B}{\partial z^2} = \frac{4 \pi^2
\beta_5(\theta)}{a^2} |B_{1x}|
\end{equation}
\begin{equation}\label{E33f}
\omega_x = \omega_{x^{\prime}} =\frac{2\pi \gamma}{a}
\sqrt{\beta_3(\theta)B_{1x}}
\end{equation}
\begin{equation}\label{E33g}
\omega_y= \omega_{y^{\prime}}=\frac{2\pi \gamma}{a}
\sqrt{\beta_4(\theta)B_{1x}}
\end{equation}
\begin{equation}\label{E33h}
\omega_z = \frac{2\pi \gamma}{a}
\sqrt{\beta_5(\theta)B_{1x}}
\end{equation}
\begin{equation}\label{E33i}
\Delta B^x = \Delta B^{x^{\prime}} =\beta_6(\theta) |B_{1x}|
\end{equation}
\begin{equation}\label{E33j}
\Delta B^z=\beta_7(\theta)|B_{1x}|
\end{equation}
\endnumparts
where $\gamma = \sqrt{{ m_{{}_F}g_{{}_F}{\mu_{{}_B}}}/{m}}$
 and $\beta_1(\theta)$, $\cdots$, $\beta_7(\theta)$  are dimensionless
parameters  which depend on ${c_0}=B_{0y^{\prime}}/B_{0y}$ and
$\theta$.  Using the definitions $c_1=1 - c_0^2$, $c_2=1 + c_0^2$,
$h_1(\theta)=1 + c_0\cos\theta$ and $h_2(\theta)=1 +c_0^2-2 c_0
\cos\theta$, we have \numparts
\begin{equation}\label{E34a}
\beta_0(\theta)={{c_0\sin\theta}\over{h_1(\theta)}}
\end{equation}
\begin{equation}\label{E34b}
\beta_1(\theta)= \frac{|c_1|}{{|h_1(\theta)|{[h_2(\theta)]^{1\over 2}}}}
\end{equation}
\begin{equation}\label{E34c}
\beta_2(\theta) = \frac{{h_1(\theta)h_2(\theta)}}{2{c_0\sin\theta}}
\end{equation}
\begin{equation}\label{E34d}
\beta_3(\theta)=\frac{2
c_0^2c_2\sin^4\theta}{|c_1h_1(\theta)|[h_2(\theta)]^{3\over 2}}
\end{equation}
\begin{equation}\label{E34e}
\beta_4(\theta)=\frac{ c_0^2[-8 c_0 \cos\theta + c_2(3+ \cos
2\theta)]\sin^2\theta}{|c_1h_1(\theta)|[h_2(\theta)]^{3\over 2}}
\end{equation}
\begin{equation}\label{E34f}
\beta_5(\theta)=\frac{4 c_0^2\sin^2\theta}{|c_1h_1(\theta)|[h_2(\theta)]^{1\over 2}}
\end{equation}
\begin{equation}\label{E34g}
\beta_6(\theta)=\frac{[c_2^3-2 c_0 c_1^2 \cos\theta -4
c_0^2c_2\cos 2\theta ]^{1\over 2}}{|h_1(\theta)h_2(\theta)|}-\beta_1(\theta)
\end{equation}
\begin{equation}\label{E34h}
\beta_7(\theta)=[1+\beta_{0}(\theta)^2]^{1\over 2}-\beta_1(\theta)
\end{equation}
\endnumparts
Subject to $B_{1y} = \beta_{0}(\theta)B_{1x}$, the conditions for
{\it non-zero} magnetic potential minima are \numparts
\begin{equation}\label{E35a}
B_{0{{y}}} \neq  B_{0{y}^{\prime}}
\end{equation}
\begin{equation}\label{E35b}
{\beta_2(\theta) B_{0y}}>|B_{1x}|>0
\end{equation}
\endnumparts
According to \eref{E33f}-\eref{E33h}, \eref{E34d} and \eref{E34e},
for a constant $c_0$, it is possible to change the trap frequencies
 by varying $\theta$ and/or the bias magnetic field ${\bf B_1}$.  Furthermore,
as \eref{E33i}, \eref{E33j}, \eref{E34g} and \eref{E34h} indicate,
we can adjust the magnetic field potential barriers by changing the
angle $\theta$, the bias magnetic field ${\bf B_1}$,  or both. ~\\
\\
\noindent \Tref{table3} gives the numerical and analytical inputs
and \tref{table4} gives calculated values for the relevant
quantities.  There is good agreement between the analytical results
for an infinite lattice and the numerical results which were
obtained in the central region of the lattice.
\begin{figure}[tbp]
\begin{center}
$\begin{array}{c}
\includegraphics[angle=0,width=13.2cm]{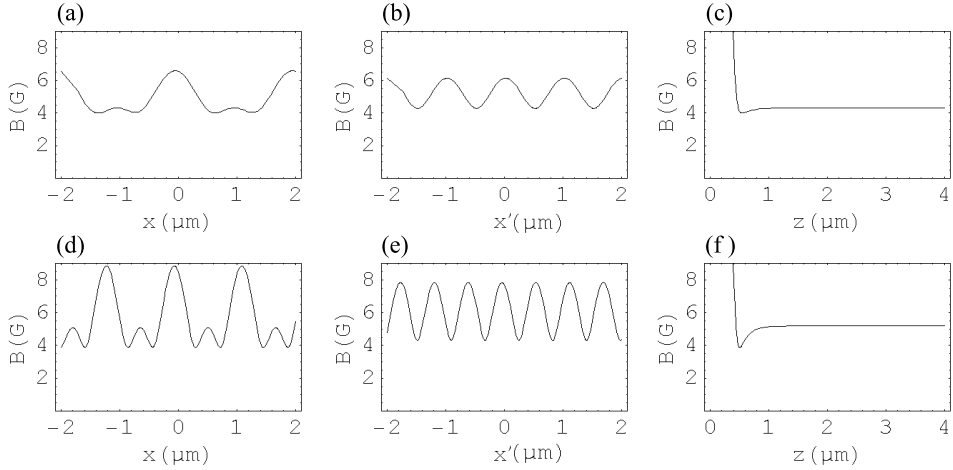}
\end{array}$
\caption{Magnetic field produced by  three arrays of parallel
rectangular magnets, as described in~\sref{Three arrays}.  The
parallel magnets in the middle and the top arrays have an angle with
respect to the parallel magnets in the lower array  of $\theta =
\pi/6$ in (a)-(c) and $\theta = \pi/3$ in (d)-(f). The curves show
the analytical magnetic field along a line ($y = y_{min}$, $z =
z_{min}$) parallel to the $x$-axis, (a) and (d), along a line
($y^{\prime}=y^{\prime}_{min}$, $z=z_{min}$) parallel to the
$x^{\prime}$-axis,  (b) and (e), and along a line ($x=x_{min}$,
$y=y_{min}$) parallel to the $z$-axis, (c) and (f).}\label{figure6}
\end{center}
\end{figure}
\section{Three periodic arrays of parallel rectangular magnets}\label{Three
arrays} Here, we consider a special configuration involving three
 periodic arrays of parallel rectangular magnets. The lower array (array 1)
and the middle array (array 2) are at an angle $\theta$ with respect
to each other, similar to the crossed arrays shown in
\fref{figure3}, while the rectangular magnets in the top array
(array 3) are parallel to those of the middle array and remain
parallel as $\theta$ changes. The period $a_3 $ is twice the periods
$a_1$ and $a_2$.  To obtain the components of the magnetic field due
to these three arrays of magnets plus bias field ${\bf
B_1}=B_{1x}{\hat x}+B_{1y}{\hat y}$, we add the magnetic field due
to  array  3 to that of arrays 1 and 2 [\eref{E21a}-\eref{E21c}]:
 \numparts
\begin{equation}\label{E36a}
\hspace{-2.5cm} B_{{x}} =  - B_{0{{y}^{\prime}}} \sin\theta \sin
(k{{y}^{\prime}}) {\rm e}^{-k{{z}}} - B_{0{{y}^{\prime\prime}}}
\sin\theta \sin (k{{y}^{\prime}}/2) {\rm e}^{-k{{z/2}}}
 + B_{{1x}}
\end{equation}
\begin{equation}\label{E36b}
\hspace{-2.5cm} B_{{y}}= [B_{0{{y}^{\prime}}}\cos\theta \sin
(k{{y}^{\prime}})+ B_{0{{y}}} \sin (k{{y}})] {\rm e}^{-k{{z}}}+
[B_{0{{y}^{\prime\prime}}}\cos\theta \sin (k{{y}^{\prime}}/2)] {\rm
e}^{-k{{z/2}}}+ B_{{1y}}
\end{equation}
\begin{equation}\label{E36c}
\hspace{-2.5cm} B_{{z}} = [B_{0{{y}^{\prime}}}  \cos
(k{{y}^{\prime}}) +B_{0{{y}}} \cos (k{{y}}) ]{\rm e}^{-k{{z}}} +
B_{0{{y}^{\prime\prime}}}  \cos (k{{y}^{\prime}}/2){\rm
e}^{-k{{z/2}}}
\end{equation}
\endnumparts
where $z \gg {a\over{\pi}}+ s_3 + t_3$ and $B_{0{{y}}}= B_{0} ( {\rm
e}^{kt_1}-1){\rm e}^{ks_1}$, $ B_{0{y}^{\prime}} = B_{0} ( {\rm
e}^{kt_2}-1){\rm e}^{ks_2}$ and $ B_{0{y}^{\prime\prime}} = B_{0} (
{\rm e}^{kt_3/2}-1){\rm e}^{ks_3/2}$. This magnetic field is shown
for $\theta = \pi/6$ and $\theta = \pi/3$ in \fref{figure6} (a)-(c)
and \fref{figure6} (d)-(f), respectively. The period $a_3 = 1$ $\mu
m$, which is twice the periods $a_1$ and $a_2$. The thicknesses of
the layers are $t_1=50$ $nm$, $t_2=40$ $nm$ and $t_3=5$ $nm$ and the
lower surfaces of the layers are at distances $s_1=0$, $s_2=60$ $nm$
and $s_3=105$ $nm$ from the plane $z=0$. The magnetization is as
before and the components of the bias magnetic field are
$B_{1x}=-4.08$ $G$, $B_{1y}=-1.39$ $G$ and $B_{1z}=0$. According to
\fref{figure6}, we can change the separation and also the barrier
height between  adjacent microtraps by varying $\theta$.  This might
be of interest for a two-qubit quantum gate process~\cite{Barenco}.
\section{Discussion and summary}
We have introduced a class of permanent magnetic lattices for
ultracold atoms and quantum degenerate gases. The potential barriers
between the adjacent microtraps can be altered  by varying the bias
magnetic field, as indicated by the relevant equations.   Thus,
these configurations of permanent magnets together with an
adjustable
 bias magnetic field could be used to study quantum tunnelling and to realize a BEC to Mott insulator
quantum phase transition for quantum information processing in a
magnetic lattice.\\

\noindent We have shown that for a configuration of two crossed
periodic arrays of
 parallel rectangular magnets with a variable angle between the
two arrays,  the magnetic lattice goes from a 1D to a 2D lattice as
the angle $\theta$ between the arrays is changed from $0$ to $\pi/
2$.  This may prove useful for loading ultracold atoms into  a 2D
lattice of 3D microtraps by initially loading the atoms into the 1D
lattice of 2D microtraps and then  by changing the angle from
$\theta = 0$ to  $\theta = \pi/2$.  Also, by varying the angle
$\theta$, it is possible to change the depth of the traps
 adiabatically,  which suggests the possibility of realizing a
`mechanical' BEC to Mott insulator quantum phase transition  in a
magnetic lattice. Moreover, the period of the magnetic lattice can
be  adjusted  as the angle $\theta$ is changed. This could  have
applications in quantum information processing  and also in creating
new artificial crystalline materials.  In a more general case, it is
possible to have two different periods $a_y = a_1$ and
$a_{y^{\prime}} = a_2$ in the $y$- and $y^{\prime}$-directions,
respectively, to allow for a more flexible magnetic lattice. A
configuration of magnets with three periodic arrays of rectangular
magnets having two different periods $a_1 = a_2 $ and $a_3$ may be
useful for  implementing  a two-qubit quantum gate~\cite{Barenco} in
a magnetic lattice.
 \ack{
This project is supported by Swinburne University strategic
initiative funding. Saeed Ghanbari would like to thank the Iranian
Government for financial support. }

\end{document}